# Kolmogorov Similarity Hypotheses for Scalar Fields: Sampling Intermittent Turbulent Mixing in the Ocean and Galaxy

Carl H. Gibson

*Departments of Applied Mechanics and Engineering Sciences
and Scripps Institution of Oceanography
University of California at San Diego, R-010
La Jolla, CA 92093-0411*

**Abstract**

Kolmogorov's three universal similarity hypothesis are extrapolated to describe scalar fields like temperature mixed by turbulence. The analogous first and second hypotheses for scalars include the effects of Prandtl number and rate-of-strain mixing. Application of velocity and scalar similarity hypotheses to the ocean must take into account the damping of active turbulence by density stratification and the earth's rotation to form fossil turbulence. By the analogous Kolmogorov third hypothesis for scalars, temperature dissipation rates averaged over lengths $r > L_K$ should be lognormally distributed with intermittency factors $\sigma^2$ that increase with increasing turbulence energy length scales $L_O$ as $\sigma^2_{\ln(\chi)_r} \mu \ln(L_O/r)$. Tests of Kolmogorovian velocity and scalar universal similarity hypotheses for very large ranges of turbulence length and time scales are provided by data from the ocean and the galactic interstellar medium. These ranges are from 1 to 9 decades in the ocean, and over 12 decades in the interstellar medium. The universal constant for turbulent mixing intermittency $\mu$ is estimated from oceanic data to be 0.44±0.01, which is remarkably close to estimates for Kolmogorov's turbulence intermittency constant $\mu$ of 0.45±0.05 from galactic as well as atmospheric data. Extreme intermittency complicates the oceanic sampling problem, and may lead to quantitative and qualitative undersampling errors in estimates of mean oceanic dissipation rates and fluxes. Intermittency of turbulence and mixing in the interstellar medium may be a factor in the formation of stars.

## 1. Introduction

Natural flows at very high Reynolds, Froude and Rossby numbers in the ocean, atmosphere, stars and the interstellar medium develop highly intermittent turbulence and mixing;

that is, viscous and scalar dissipation rates become more and more nonuniform in space and time as the range of scales of the turbulence increases, despite the strong tendency of turbulence to mix and homogenize these quantities. Note that here the term *intermittency* indicates a wide tail in a skewed probability density function for the viscous or scalar dissipation rates. It is not the fraction of turbulence versus nonturbulence, or mixing versus nonmixing, as sometimes defined. Estimates of vertical momentum fluxes in the ocean and atmosphere by dissipation techniques[1] require the inference of average viscous dissipation rates from measurements of local values. Large intermittency increases the danger of undersampling errors. Kolmogorov's three universal similarity hypotheses provide powerful tools for extracting a large amount of information about turbulence from a small number of samples. Kolmogorov similarity is particularly valuable for natural flows because these are generally vastly undersampled. It should be possible, and useful, to extrapolate the Kolmogorov theory to describe scalar fields like temperature mixed by turbulence. Both turbulence and turbulent mixing parameters may be inferred from sparse samples of scalar fields using analogous universal similarity hypotheses for turbulent mixing. Scalars are often easier to measure than the velocity, and form fossil turbulence[2] features that are generally more robust than fossil turbulence velocity fields in preserving information about previous turbulence. Because turbulence in natural flows is often constrained at large scales to form such relics, fossilization effects on similarity must be understood in the interpretation of data.

The present paper extrapolates Kolmogorov universal similarity hypotheses to scalar fields mixed by turbulence in natural flows, and provides some comparisons to data. The extrapolation is complicated by effects of Prandtl, Froude and Rossby numbers, but is generally quite feasible and successful. The Prandtl number determines the smallest scale fluctuations of scalar fields mixed by turbulence, and Froude and Rossby numbers of the flow determine the largest turbulence scales possible before buoyancy or Coriolis forces constrain the turbulent inertial-vortex forces. Fluctuations may persist as fossil turbulence in scalar fields after the fluid ceases to be actively turbulent at the scale of the fluctuation. These remnants provide information about the previous turbulence, and are most useful in avoiding undersampling errors that might occur if the dominant turbulent events in a flow are damped so rapidly that they are difficult to detect by sparse sampling.

The largest ranges of turbulence scales accessible by direct measurements, from diffusive to energy scales, are presently in the ocean. Temperature sensors on airplanes lack the necessary frequency response to measure diffusive scale temperature fluctuations in the atmosphere.

---

[1]Steady state is assumed for a control volume layer perpendicular to a turbulent momentum flux, so the production of turbulent kinetic energy (the product of flux and mean velocity gradient) is equal to the viscous dissipation rate.

[2]Fossil turbulence is a fluctuation in a scalar or flow field produced by turbulence that persists after the fluid is no longer actively turbulent at the scale of the fluctuation. Most ocean microstructure is fossil turbulence.

However, microconductivity sensors mounted on high speed towed bodies have no frequency response limitation, and are able to resolve the millimeter to megameter length scale range required for temperature and velocity fluctuations in the ocean. Monin and Ozmidov (1985) describe the use of such sensors and platforms throughout the world's oceans. Kolmogorov himself participated in such cruises, as noted by Kendall (1990), along with his students Oboukhov, Monin and Ozmidov.

## 2. Universal Similarity Hypotheses for Scalar Fields

The power of the Kolmogorov (1941, 1962) three universal similarity hypotheses for high Reynolds number turbulent velocity fields is in their universality. By hypothesis, turbulence[3] is turbulence, everywhere in the universe. It seems natural to assume that many aspects of turbulent mixing are also universally similar, and that the Kolmogorov hypotheses for turbulence have parallel counterparts for turbulent mixing. Oboukhov (1949) and Corrsin (1951) made these assumptions and implicitly extended the second Kolmogorov hypothesis to scalar mixing in their independent predictions of scalar inertial subranges in structure functions and spectra

$$\Phi = \beta_{si} \chi \varepsilon^{-1/3} k^{-5/3} \; ; \; L_O^{-1} < k < L_C^{-1} \; ; \; L_C \equiv \left(\frac{D^3}{\varepsilon}\right)^{1/4} \tag{1}$$

where $\Phi$ is the spectrum of a conserved, dynamically passive, scalar fluid property $\theta$ like temperature, with integral $\overline{\theta^2}$, $\chi$ is the dissipation rate of scalar variance, $\varepsilon$ is the viscous dissipation rate, k is the radian wavenumber magnitude, $L_O$ is the energy, or Oboukhov, scale, $L_C$ is the Oboukhov-Corrsin inertial-diffusive scale, D is the molecular diffusivity of $\theta$, and $\beta_{si}$ is a universal constant of order one. Equation (1) is parallel to equation (2) for the inertial subrange of turbulence that follows from the second Kolmogorov hypothesis

$$\Phi_u = \alpha \varepsilon^{2/3} k^{-5/3} \; ; \; L_O^{-1} < k < L_K^{-1} \; ; \; L_K \equiv \left(\frac{\nu^3}{\varepsilon}\right)^{1/4} \tag{2}$$

where $\Phi_u$ is the spectrum of the velocity component u with integral $\overline{u^2}/2$, $L_K$ is the Kolmogorov inertial-viscous scale, $\nu$ is the kinematic viscosity, and $\alpha$ is a universal constant of order one.

Batchelor (1959) realized that the direct Oboukhov-Corrsin extrapolation of (2) to derive (1) ignores important physical mechanisms of turbulent mixing that are Prandtl number dependent. For Pr $\equiv \nu/D > 1$ the smallest scalar fluctuations would be on scales $L_C < L_K$ according to (1), so that such fluctuations would experience a velocity field of uniform straining at a rate $(\varepsilon/\nu)^{1/2}$. From a Fourier model for weakly diffusive scalars with $Pr > 1$, Batchelor (1959) showed the rate-of-strain should be the relevant dimensional parameter for the smallest scale mixing process, rather than $\varepsilon$ as assumed in the derivation of (1). He derived

$$\phi = \beta_{vc} \chi \gamma^{-1} k^{-1} \exp\left(-\beta_{vc} k^2 D/\gamma\right) ; L_K^{-1} > k \tag{3}$$

where $\beta_{vc}$ is a universal constant for the viscous-convective subrange; that is, for scales smaller than the Kolmogorov scale $L_K$. The spectrum in (3) has an exponential diffusive cutoff at the Batchelor length scale $L_B \equiv (D/\gamma)^{1/2}$. Several experimental studies have confirmed (3) in the laboratory, eg. Gibson and Schwarz (1963b), and ocean, eg. Dillon and Caldwell (1980). For $Pr < 1$ the Batchelor Fourier model fails because the wavelength of the smallest scalar fluctuation is larger than the scales of uniform straining. Consequently, Batchelor, Howells and Townsend (1959) and others have concluded that $\gamma$ should be irrelevant to the mixing of scalars with $Pr < 1$.

---

[3]Turbulence is defined as an eddy-like state of fluid motion where the inertial-vortex forces of the eddies are larger than the viscous, buoyancy, Coriolis, electromagnetic, or any other forces which tend to damp the eddies.

However, Gibson (1968a) showed that non-Fourier $\gamma$-mixing mechanisms exist that are independent of Pr because they involve straining of scalar topological features, such as ridge lines and extremum points, that are always smaller than $L_K$. Therefore, Gibson (1968b) proposed a set of universal scalar similarity hypotheses, parallel to those of Kolmogorov (1941), that assume $\gamma$ determines the smallest scales for all Pr. Gibson (1981) extended the Kolmogorov (1962) refinement to account for intermittency effects on the universal hypotheses for scalar fields. The proposed scalar universal similarity hypotheses 1ab, 2ab, 3ab and 4 for turbulent mixing are compared to the Kolmogorov hypotheses 1, 2 and 3 for turbulence in Table 1.

**Table 1. Universal Similarity Hypotheses of Turbulence and Turbulent Mixing**

Turbulence

| Hypothesis | Length range |
|---|---|
| 1. $F_n(\varepsilon, \nu, y_k)$ | $y_k < L_O$ |
| 2. $F_n(\varepsilon, y_k)$ | $L_K < y_k < L_O$ |
| 3. $\overline{\ln(\varepsilon)^2}_r = A + \mu \ln\left(\frac{L_O}{r}\right)$ | $L_K < r < L_O$ |

Turbulent Mixing

| Hypothesis | Length range | Prandtl number, Pr |
|---|---|---|
| 1a. $F_n(\ ,\ , D, y_k)$ | $y_k < L_B$ | all values |
|  | $y_k < L_K$ | $> 1$ |
|  | $y_k < L_C$ | $< 1$ |
| 1b. $F_n(\ ,\ , y_k)$ | $L_K < y_k < L_O$ | $> 1$ |
|  | $L_C < y_k < L_O$ | $< 1$ |
| 2a. $F_n(\ ,\ , D, y_k)$ | $L_B < y_k < L_O$ | $< 1$ |
| 2b. $F_n(\ , D, y_k)$ | $L_B < y_k < L_C$ | $< 1$ |
| 3a. $F_n(\ ,\ ,\ , y_k)$ | $L_B < y_k < L_O$ | $> 1$ |
| 3b. $F_n(\ ,\ , y_k)$ | $L_B < y_k < L_K$ | $> 1$ |
| 4. $\ln^2 \langle\ \rangle_r = A + \mu \ln\left(\dfrac{L_O}{r}\right)$ | $L_K < r < L_O$ | $> 1$ |
|  | $L_C < r < L_O$ | $< 1$ |

Kolmogorov's first similarity hypothesis is that the three-n joint distribution function $F_n$ for velocity differences between points in high Reynolds number turbulence should be locally isotropic in space and time, and depend only on  ,   and the magnitudes $y_k$ of the n vectors separating the points; that is, $F_n = f(\ ,\ , y_k)$ for $y_k < L_O$. Length and time transformation to values normalized by the Kolmogorov length $L_K$ and the Kolmogorov time $T_K$   $(\ /\ )^{1/2} =\ ^{-1}$ reduces the probability laws to universal forms.  Therefore, all statistical parameters such as Kolmogorov normalized spectra and structure functions should be universal.

The analogous first similarity hypothesis in Table 1 for scalar mixing is 1a; that the n-joint distribution function $F_n$ for scalar differences between the n points should be locally isotropic in space and time, and depend only on  ,  , D and $y_k$ for $y_k < L_K$ (Pr $> 1$) or  $y_k < L_C$ (Pr $< 1$), and for $y_k < L_B$ for all Pr.  Thus length, time and scalar transformations to values normalized by the Batchelor length $L_B$, time $T_B =\ ^{-1}$ and scalar scale $S_B$   $(\ /\ )^{1/2}$ reduces the probability laws and all derived statistical parameters to universal forms. Measurements of temperature fluctuations in turbulent mercury, Pr = 0.02, by Clay (1973) and numerical simulations by Kerr (1985, 1990) and Gibson, Kerstein and Ashurst (1988) confirm the predicted universal Batchelor similarity for turbulent mixing with Pr values from 0.02 to 10.  Van Atta (1991, this volume) presents evidence supporting the assumptions of local isotropy for turbulent velocity and scalar fields.for a variety of laboratory and natural flows.

The scalar analog of Kolmogorov's second hypothesis in Table 1 is 1b; that $F_n(\ ,\ , y_k)$ for $L_K < y_k < L_O$ and Pr $> 1$, and $L_C < y_k < L_O$ for Pr $< 1$. Hypothesis 1b predicts the $k^{-5/3}$ scalar inertial subrange in (1) by dimensional analysis, but the smallest scales of the subranges are $L_K$ or $L_C$ for large and small Pr, respectively. Similarity hypotheses 2ab and 3ab in Table 1 have no direct equivalents in Kolmogorov's hypotheses, and depend on the mixing mechanisms assumed. Hypothesis 2b predicts a $k^{-3}$ spectral subrange for Pr < 1, and hypothesis 3b predicts a $k^{-1}$ subrange for Pr > 1. Hypothesis 2a indicates convergence of the probability laws for Pr < 1 in scales normalized by Corrsin length $L_C$, time $T_C$ ( $Pr^{1/2})^{-1}$ and scalar $S_C$ ( $T_C)^{1/2}$ scales. Hypothesis 3b indicates convergence of the probability laws for Pr > 1 in scales normalized by Kolmogorov length $L_K$, time $T_K$ and scalar $S_K$ ( $T_K)^{1/2}$ scales. Similarity hypothesis 4 for scalar mixing is an extrapolation of Kolmogorov's hypothesis 3 for turbulence based on the Gurvich and Yaglom (1967) cascade model, applied by Gibson (1981) to oceanic mixing.

## 2. Kolmogorov Similarity Hypotheses in the Galaxy

Since the Kolmogorov (1941) theory is for asymptotically large Reynolds number turbulence, it seems appropriate to compare with data with a maximum range of scales. Figure 1 shows a comparison between hypothesis 1b of Table 1 and the three dimensional spectrum $P_{3N}$ of the electron density fluctuations of the local interstellar medium, from Armstrong, Cordes and Rickett (1981). The spectrum gives the variance $\overline{\phantom{xx}}^2$ when integrated over three components of wave number of magnitude q rad m$^{-1}$, so the expected logarithmic slope is -11/3, shown by the dark line of Fig. 1 through the data.

As can be seen from Fig. 1, the electron density fluctuation spectrum is in good agreement with the scalar inertial subrange hypothesis 1b of Table 1 over an enormous range of scales: nearly 12 decades of wavelength, from $10^{19}$ to $10^7$ meters; that is, from the thickness of the galaxy to the size of the earth. A wide variety of questions might be raised regarding the interpretation of this apparent agreement as any confirmation of Kolmogorov similarity, stemming from uncertainties about the fluid mechanics of the interstellar medium and whether electron density can be considered a passive scalar additive mixed by turbulence for the various data sources represented. Nevertheless, the agreement is intriguing, and the possible connection to turbulent scalar similarity theory seems worth pursuing.

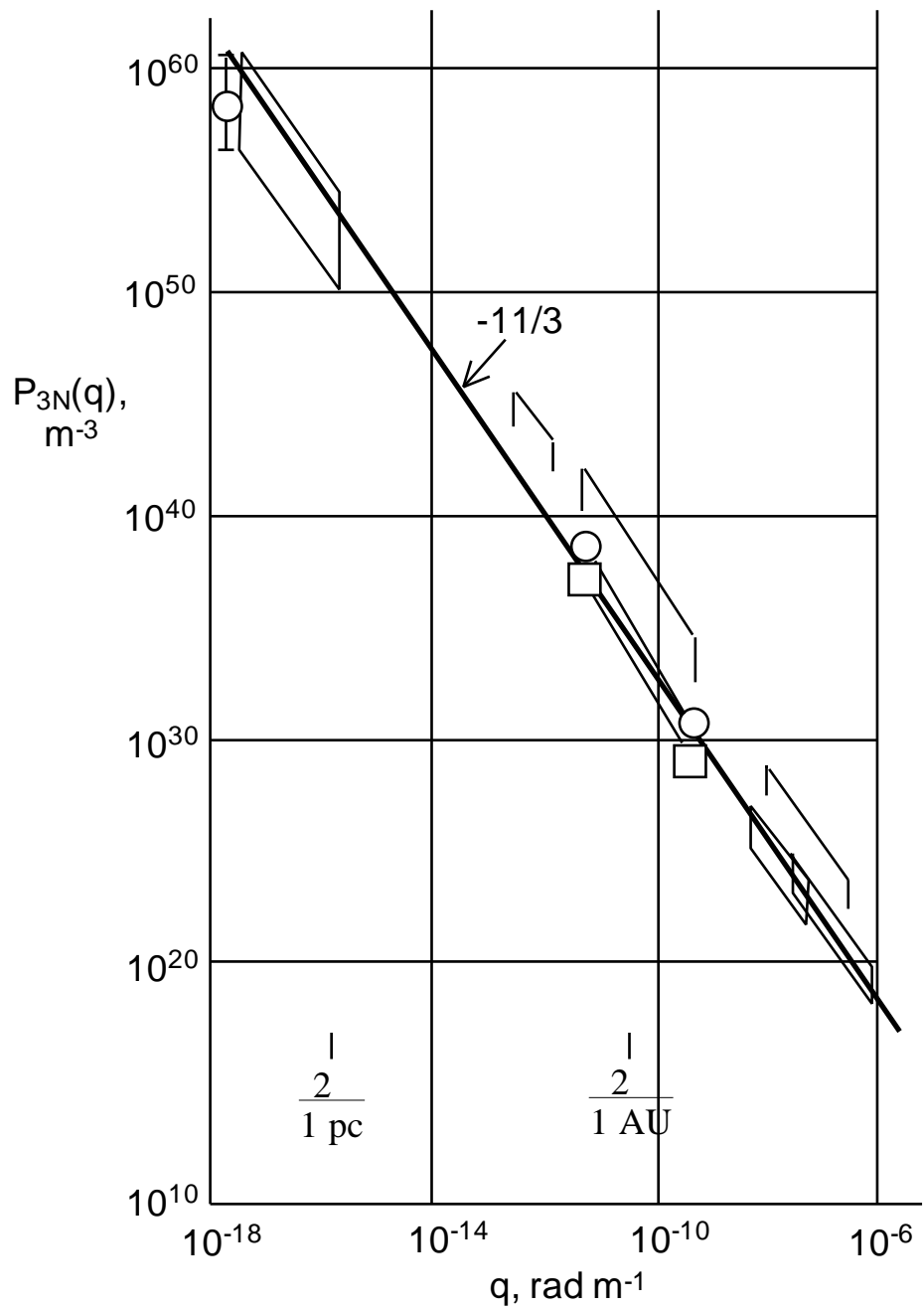

**Figure 1.** Electron density power spectrum in the local interstellar medium. Estimates and limits come from a variety of astronomical sources.

Figure 2 shows a comparison of Kolmogorov's second hypothesis for turbulence with the velocity difference (magnitude) structure function for length scales $10^{15}$ to $10^{19}$ m, inferred from line broadening of molecular emission spectra from dense molecular interstellar gas clouds in this and other galaxies by Falgarone and Phillips (1990). Dense means about $10^4$ atoms per cubic centimeter. The gas is very cold, less than 10 K, but the emission lines are broad and non-Gaussian in form, suggesting doppler shifts due to intermittent turbulence.

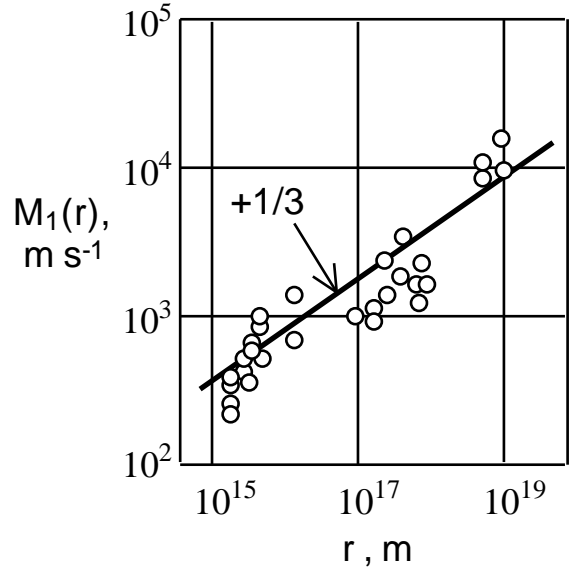

**Figure 2.** First moment $M_1$ of the velocity difference magnitude as a function of the separation distance r between points in interstellar gas clouds, Falgarone and Phillips (1990).

From the second Kolmogorov hypothesis of Table 1, it is possible to estimate the dissipation rate of the turbulence implied by Fig. 2. According to this hypothesis,

$$M_n \equiv \langle |\vec{v}(\vec{x}+\vec{r})-\vec{v}(\vec{x})|^n \rangle \approx (\bar{\varepsilon}_r r)^{n/3} r^{-\mu n(n-3)/18} \qquad (4)$$

where the dissipation rate averaged over control volumes of size r is assumed to be a lognormal random variable, according to hypothesis 3 of Table 1. From higher order velocity difference moments and (4), a constant $\mu \approx$ 0.4-0.5 is indicated for the third hypothesis of Kolmogorov, Table 1. This is in good agreement with experimental values from marine atmospheric boundary layer measurements, discussed in §5.

Taking the proportionality constant of (4) to be of order one gives $\bar{\varepsilon} \approx$ 7x10$^{-8}$ m$^2$ s$^{-3}$. For comparison, this is a typical value for the dissipation rate $\varepsilon$ in the upper few meters of the ocean. Using this value in

$$P_{3N} \approx \varepsilon^{-1/3} q^{-11/3} \qquad (5)$$

and taking the proportionality constant to be one, gives a dissipation rate for the electron density variance of     $4.1 \times 10^{-7}$ m$^{-6}$s$^{-1}$.

The lifetime of interstellar molecular cloud complexes is estimated by Scheffler and Elsässer (1987) to be 30 to 50 times longer than that expected from purely gravitational collapse because the condensation of the gas to form stars is inhibited by turbulence. The extreme intermittency of the turbulence dissipation rate , as well as its magnitude and geometry, will clearly affect the probability of star nucleation. As discussed by Gibson (1988), nucleation occurs if the gravitational scales $L_g$   GM/  or $L_{g'}$   (GM)$^{3/5}$/ $^{2/5}$ become larger than a region of mass concentration, where G is the gravitational constant and M is the mass. Buoyancy forces will arise that damp the turbulence and accelerate the self gravitational condensation process when $L_R$ decreases to equal the increasing $L_g$ or $L_{g'}$. The resulting solar system is a form of fossil turbulence because the size preserves information about previous turbulence.

## 3. Kolmogorov Similarity in the Ocean

The first strong confirming evidence for Kolmogorov's first and second universal similarity hypotheses over a wide range of Reynolds numbers was provided by measurements with hot film anemometers on a towed body in an oceanic tidal channel by Grant, Stewart and Moilliet (1961), compared to grid turbulence measurements of Stewart and Townsend (1951) in air and Gibson and Schwarz (1963b) in water, as shown in Figure 3.

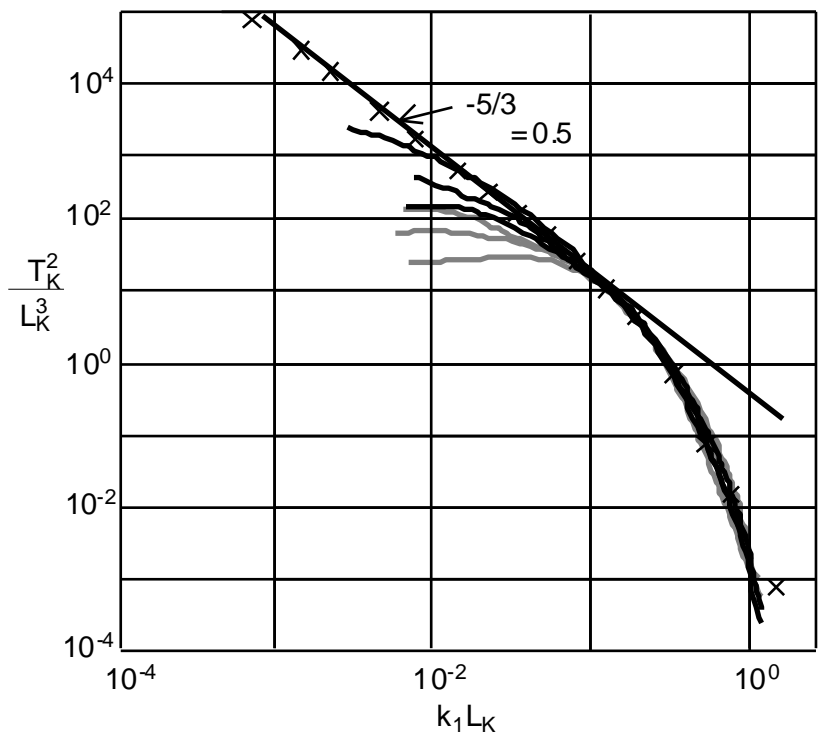

**Figure 3.** Turbulence velocity spectra from an oceanic tidal channel, X's, water tunnel grid wake, dark line, and wind tunnel grid wake, light line, normalized with Kolmogorov length and time scales, from Gibson and Schwarz (1963b).

The spectra for all the flows in Fig. 3 converge at high wavenumbers, as predicted by Kolmogorov's first similarity hypothesis, and show good agreement with the same tangential curve with logarithmic slope -5/3 predicted by the Kolmogorov second similarity hypothesis. The indicated universal inertial subrange constant   from (2) is about 0.5.

Velocity and temperature microstructure from breaking internal waves caused by tidal flows over a sill in Knight Inlet were measured from a submersible far downstream by Gargett, Osborn and Nasmyth (1984). The velocity spectra, multiplied by $k^{-5/3}$ and normalized by Kolmogorov length and time scales, show good agreement with the universal spectrum, as shown in Figure 4.

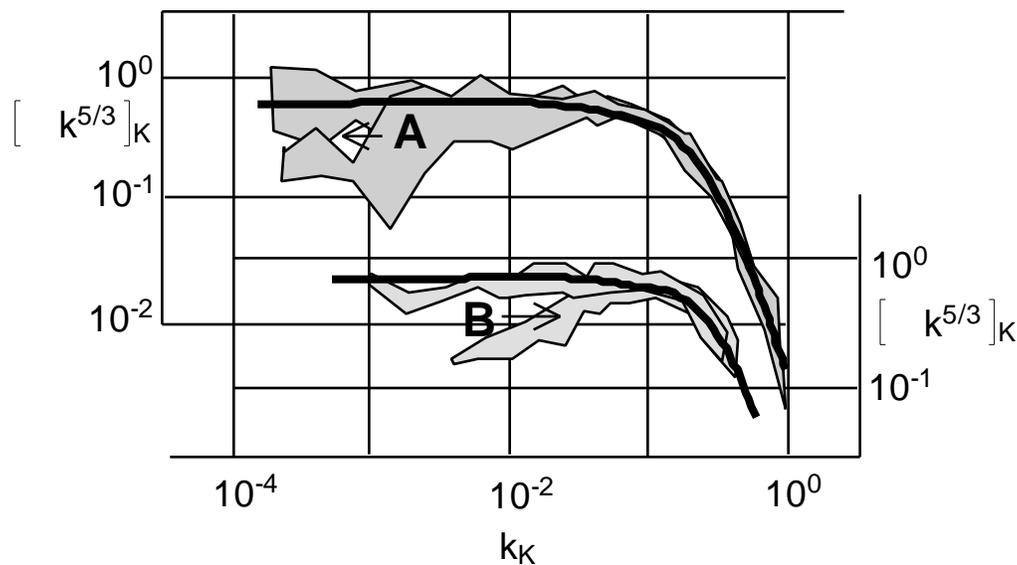

**Figure 4.** Comparison of the Kolmogorov normalized universal spectral form for velocity spectra of Fig. 3 times $k^{-5/3}$, dark lines, with measurements in Knight Inlet from a submersible, by Gargett, Osborn and Nasmyth (1984). Spectra from the most active regions (A), left ordinate, are outlined with hash marks. Spectra from quieter regions (B), right ordinate shifted down, are outlined with dots. All measured spectra agree with the universal form implied by the data of Fig. 3.

However, Gargett (1985) found large discrepancies between temperature spectra measured in regions A and B and the universal spectral form expected from the Batchelor (1959) theory and the laboratory measurements of Gibson and Schwarz (1963b). The universal spectrum for Pr = 10 is shown by the dark lines in Figure 5 compared to data envelopes. Temperature spectra were computed from the same strong and weak activity regions of Fig. 3, multiplied by $k^{-5/3}$ for comparison to a scalar inertial subrange, and normalized by Kolmogorov scalar scales $L_K$, $T_K$, and scalar scale $S_K$ ( $T_K)^{1/2}$. As shown in Fig. 5, the largest discrepancies are about a factor of 4 for the strong activity regions (A). Why should the velocity field for these most active regions agree with universal similarity, but the scalar fields from the same regions so strongly disagree?

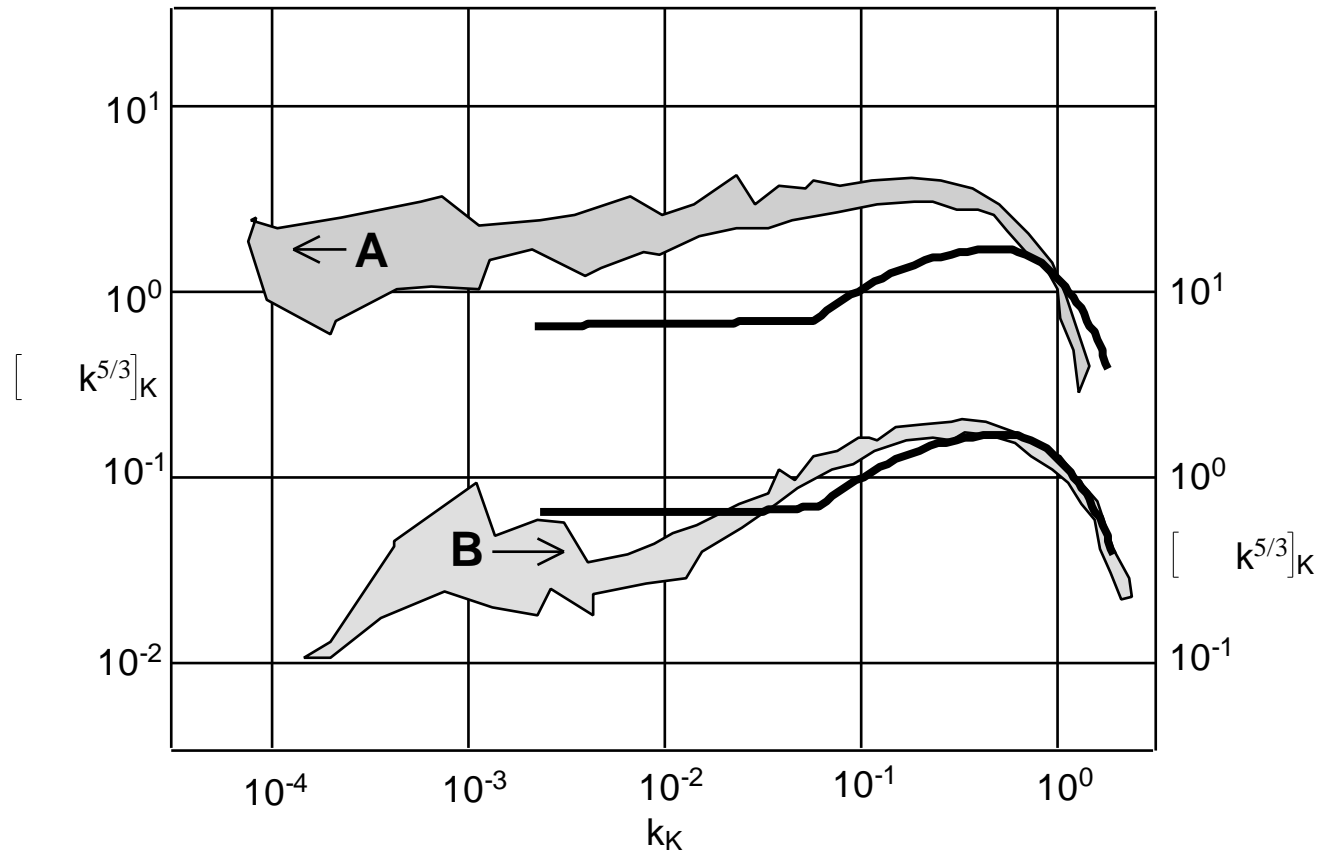

**Figure 5.**  Apparent failure of scalar universal similarity for temperature fluctuations in Knight Inlet, from Gargett (1985). The universal spectral form, in Kolmogorov coordinates for Pr = 10, is show by dark lines compared to data envelopes for strong (A: left ordinate) and weak (B: right ordinate shifted down) activity regions, using the same data envelope shading as in Fig.4.  The disagreement may be due to fossil turbulence effects.

The failure of the temperature microstructure to obey universal similarity scaling in Fig. 5 is attributed by Gargett (1985) to a failure of universal similarity theory for scalars. Gibson (1987) proposed an alternative explanation; that is, that the strong stratification of the fjord converts the larger scales of the turbulence to saturated internal waves, producing patches of fossil temperature turbulence with persistent large values but with for the embedded turbulence much smaller than existing when buoyancy forces first affected the original patch. Gargett (1985) normalized the spectral estimates in Fig. 5 using the bulk average $_{bulk}$ for the entire file, and averaged together spectra computed from 10 m records, longer than the local turbulence scales. However, the similarity hypotheses of Table 1 require that record lengths for averaging be no longer than the maximum turbulence scale $L_O$ for the record, so the hypotheses may not be obeyed in Figs. 4 or 5.

In a stable stratified fluid, the maximum turbulence scale is limited by the Ozmidov scale $L_R$ ( $/N^3)^{1/2}$, where N [g( / z)/ ]$^{1/2}$ is the stratification frequency, g is gravity, is density, and z is depth. For active turbulence, $L_R \gg L_O$. When decreases for decaying turbulence in a stratified fluid, $L_R$ may decrease to match $L_O$. At this point fossilization begins: the turbulence eddies are converted to internal waves, leaving persistent microstructure in the temperature and salinity fields known as fossil temperature turbulence and fossil salinity turbulence, as described by Gibson (1980, 1986). Within such fossils, is large but is small. This is the distinguishing signature of fossil turbulence. Consider the expressions for the Kolmogorov normalization of turbulent velocity and scalar spectra

$$\text{velocity:} \langle k^{5/3} {}_u \rangle_K = \langle k^{5/3} {}_u {}^{-2/3} \rangle$$
$$\text{scalar:} \langle k^{5/3} \rangle_K = \langle k^{5/3} {}^{-1} {}^{+1/3} \rangle \tag{6}$$

where the dissipation rates and spectral levels averaged on the right of (6) must first be averaged over scales smaller than the maximum local turbulence scale by the hypotheses of Table 1 in order for their weighted products to be universal, so that their average can converge to the universal average on the left.

What is the effect of stratification forming fossil turbulence? For the velocity field, the fossil patches will have spectral levels $_u$ and $_{fossil}$ values smaller than the bulk averages. Their contribution to the average Kolmogorov normalized spectrum will be further suppressed because the spectral weighting factor $^{-2/3}$ in (6) has a negative power and $_{bulk} \gg {}_{fossil}$. The estimate of the normalized spectrum formed by bulk averaging will be dominated by the regions in the record of maximum spectral level and maximum that are fully turbulent. Therefore it will tend toward the universal curve, as in Fig. 4, or be slightly decreased because the nonturbulent portions of the record have not been deleted. In contrast, the average Kolmogorov normalized scalar spectra are increased rather than decreased by fossil scalar turbulence patches using the Gargett (1985) averaging procedure because fossil contributions have positive weighting factors approximately equal to ( $_{bulk}$/ $_{fossil})^{1/3}$ from (6). These weighting factors can be several orders of magnitude in a stratified field of active and fossil turbulence in the ocean, which could account for the factor of 4 departure of the bulk normalized spectra in Fig. 5 from the universal curve for the regions of strong activity, and smaller departures for the regions of weak activity where the weighting factors will be smaller. Further discussion of these results are given by Phillips (1991) in this volume.

## 4. Undersampling Errors in the Ocean due to Intermittency

Turbulence in the ocean and atmosphere is extremely intermittent compared to turbulence in the laboratory. From several ocean microstructure data sets, Baker and Gibson (1987) estimate the intermittency factors $\sigma^2_{\ln \varepsilon}$ and $\sigma^2_{\ln \chi}$ for most internal layers to be between 3 and 7, versus 1-2 for the surface mixed layer or for most laboratory turbulence. Dissipation rates $\varepsilon$ and $\chi$ have measured distribution functions that are highly non-normal, and indistinguishable from lognormal, as expected from the Kolmogorov third hypothesis for turbulence, and from the scalar version of this hypothesis for turbulent mixing in Table 1. Such enormous intermittency factors imply that the turbulence and mixing in such layers are dominated by horizontal turbulence scales that are not limited by the maximum vertical turbulence scale $L_R$ permitted by buoyancy forces, but by the Coriolis force equivalent, the Hopfinger scale $L_H \approx (\varepsilon/f^3)^{1/2}$, which may be tens or hundreds of kilometers depending on the forcing and the latitude, where the Coriolis parameter f is $2\Omega\sin\phi$ and $\phi$ is the latitude. For example, if the intermittency factor is 5 and the Kolmogorov intermittency constants µ or µ$_\theta$ are about 0.5, this implies a range of scales of the turbulence or mixing of about 20,000. Such a range of vertical scales is not possible for the stratified ocean, where the maximum vertical overturn scale is about 10 meters. Furthermore, a very large number of independent samples must be collected to achieve satisfactory confidence intervals for such intermittent random variables. Using an expression from Baker and Gibson (1987), about 100 independent samples are needed to give a factor of two confidence interval for the lognormal maximum likelihood estimator of the mean for this example, over 1000 samples for 20% accuracy, and nearly 10,000 for 10% accuracy. Rapidly repeated sampling within a region smaller than the largest past or present energy scale of the turbulence does not increase the number of independent samples.

If the extreme intermittency of oceanic turbulence is not recognized, it can cause enormous undersampling errors in estimates of average values of the dissipation rates needed in dissipation flux estimates. The mean to mode ratio for a lognormal random variable is exp($3\sigma^2/2$), where $\sigma^2$ is the variance of ln $\varepsilon$ or ln $\chi$. If $\sigma^2$ is 5, then this ratio is 1808. Therefore, a single sample of the dissipation rate in a layer, or the average of any number of samples within the same region at the same time, is likely to be about 2000 times smaller than the mean value. Unfortunately, many investigators studying oceanic turbulence have not recognized the extreme intermittency and extreme range of length and time scales of in their data or in their sampling strategies, and have not recognized that their published microstructure is generally fossilized turbulence at the largest scales of the dominant patches, indicating it must be undersampled. A common procedure is to take samples with dropsondes that move slowly enough to permit temperature sampling to the diffusive scale by thermistors: 8-10 cm/s. Several hours are required to deploy and recover an instrument from kilometer depths so that only one or two samples a day are possible. The goal is to estimate the temperature dissipation rate normalized by the dissipation rate of the mean gradient, or Cox number,

$$C \equiv \frac{\overline{\nabla T \cdot \nabla T}}{\overline{\nabla T} \cdot \overline{\nabla T}} \tag{8}$$

because the vertical eddy diffusivity is the mean Cox number times the molecular diffusivity; that is

$$K_z \equiv D\overline{C} \tag{9}$$

from Osborn and Cox (1972), who first applied the scalar dissipation technique to estimate vertical diffusion and heat fluxes in the ocean.

Figure 6 shows a lognormal probability plot of the cumulative distribution function for a subset of the dropsonde samples of Cox numbers in the upper main thermocline of the central North Pacific, at 28°N and 155°W, collected by Gregg (1977) in three expeditions in three seasons during three years, so that the samples should be independent. In such a plot, lognormal random variables fall on straight lines with slopes $1/\sigma$ and median values at $[X-\mu]/\sigma = 0$, where X is the natural logarithm of the random variable, and $\mu$ is the mean and $\sigma$ is the standard deviation of X. The probability that a sample x is less than X, the cumulative distribution function, is shown on the right ordinate, and is estimated from the data histogram.

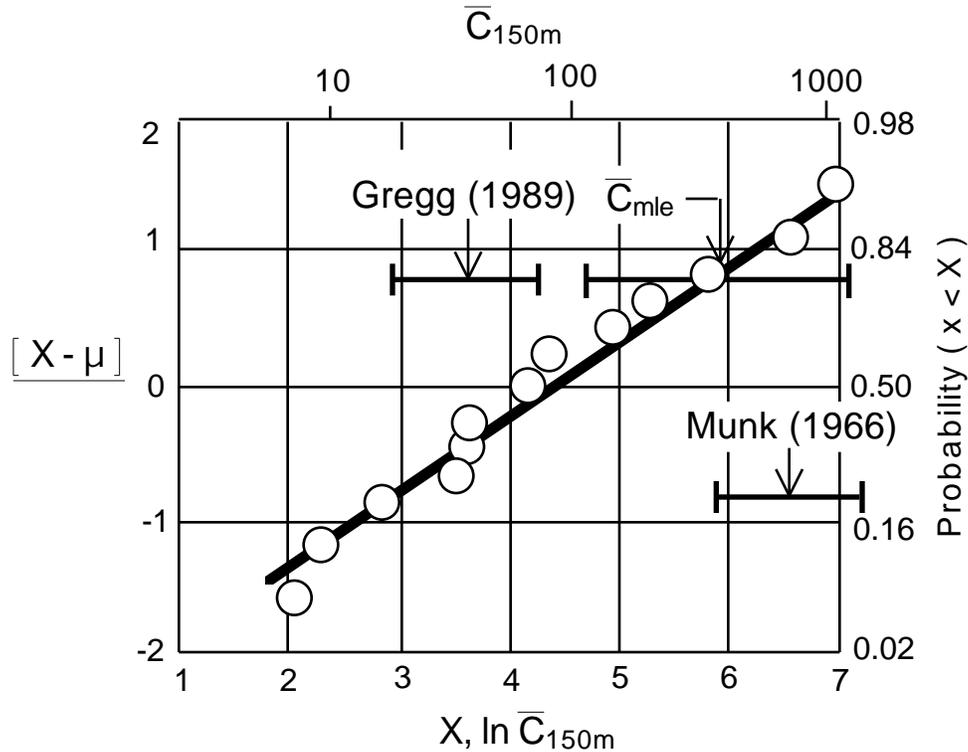

**Figure 6.** Lognormal probability plot of dropsonde Cox number samples, averaged over 150 m in the vertical, in the depth range 75-1196 m, from Gregg (1977), expedition Tasaday 11, February, 1974. The maximum likelihoodestimate of $\overline{C}$ is 379 versus 43 from the Gregg (1989) correlation: confidence intervals shown by horizontal bars do not overlap so the discrepancy is statistically significant. The $\overline{C}_{mle}$ and Munk (1966) estimates are indistinguishable.

Fig. 6 shows the Gregg (1977) Cox number samples in the upper thermocline depth range 75-1196 m fit a lognormal random variable with $\sigma^2$ of about 3.07. The maximum likelihood estimator of the mean $\overline{C}$ is 379 and the 95% confidence interval is a factor of 3, from expressions in Baker and Gibson (1987). Thus 13 of the 15 samples are less than the indicated expected value, and the mode of the distribution is 100 times less, or about 4. In the interpretation of such data, ignoring its lognormality and intermittency, values near the mode are often taken to be representative of the mean, giving a quantitative undersampling error of order 1000% and a large, but erroneous, discrepancy between the indicated vertical heat flux from microstructure measurements and values estimated from bulk flow models such as Munk (1966).

From the present Kolmogorovian interpretation of the Gregg (1977) $C_{150m}$ data as a lognormal random variable with large intermittency, the vertical heat flux down through the upper

Gibson, C. H., *Kolmogorov similarity ...in the ocean and galaxy*, Proc. Roy. Soc. Lond. A (1991) 434, 149-164thermocline of the central North Pacific decreases from about 6 watt m$^{-2}$ at a depth of 100 m to about 3 watt m$^{-2}$ at 2 km, within uncertainty factors of 3-6. From the expression for in the thermocline proposed by Gregg (1989), based on dropsonde data that indicate much smaller intermittency factors and values and which ignore the fossil turbulence evidence that these may be underestimates, the heat flux decreases from about 1 watt m$^{-2}$ at 100 m to only 0.1 watt m$^{-2}$ at 2 km, with uncertainty factors of less than 2. The Cox number from the Gregg (1989) assumption that C 0.2 /DN$^2$ is about 43 within a factor of 2, and is compared with $C_{mle}$ and its confidence interval in Fig. 6. The two estimates differ by nearly a decade. Even though the uncertainty bands are very wide, they do not overlap. Similar plots for the full data set and other subsets from Gregg (1977) lead to the same conclusions: C is lognormal with $^2$ from 3 to 7, the Gregg (1989) correlation for in the ocean thermocline is too small by a factor of 10 or more, and there is no statistically significant discrepancy between microstructure and bulk flow flux estimates of heat flux and eddy diffusivities.

Because the range of turbulent length scales in the ocean can vary considerably with depth, depending on whether the turbulence is three dimensional or two dimensional, the intermittency factors $^2$ can be strongly depth dependent. Therefore the magnitude of the expected quantitative undersampling error from dropsonde measurements varies with depth. The layers with the largest temperature (density) dissipation rates are likely to be those with maximum intermittency, and therefore maximum undersampling errors. Consequently, layers such as the seasonal thermocline which has the maximum temperature dissipation rate relative to layers above and below, will probably be identified as a layer of minimum dissipation rate by dropsonde sampling alone. Such misidentifications of layers of maximum dissipation as minima are termed qualitative undersampling errors. For further discussion see Gibson (1990).

## 5. Lognormal Intermittency of Temperature Dissipation in the Upper Ocean

Soviet oceanographers, under the leadership of Monin for over twenty years, have adopted a different strategy for sampling oceanic turbulence and mixing. High speed towed bodies with high frequency response, high spatial resolution microconductivity probes have been used to detect temperature fluctuations to the diffusive scale. With electrical current densities high enough to cause heating, conductivity probes become hydroresistance anemometers, and have been used to sample velocity fluctuations to the the viscous scale. Using a large fleet of large ships, a wide variety of oceanic layers and flows have been sampled in all the world's oceans. Extremely intermittent turbulence and turbulent mixing, as well as Kolmogorov turbulence and mixing similarity, have been clearly demonstrated in these measurements; for example, by Monin and Ozmidov (1985) and Belyaev et al. (1974). Spatial resolution and laboratory applications of similar microconductivity probes are discussed by Gibson and Schwarz (1963a).

Figure 7 shows a comparison to normality and lognormality of 4 m averaged values measured on a towed body in a California current upwelling region off Monterey Bay, during the 51st Cruise of the AKADEMIK KURCHATOV, September 1990. Data were kindly provided by Chief Scientist Lozovatskiy to the authors, who participated in the Hilo to Brisbane leg of the cruise. The data analysis was carried out by Dr. Baker. A detailed discussion of these preliminary results will be published in due course.

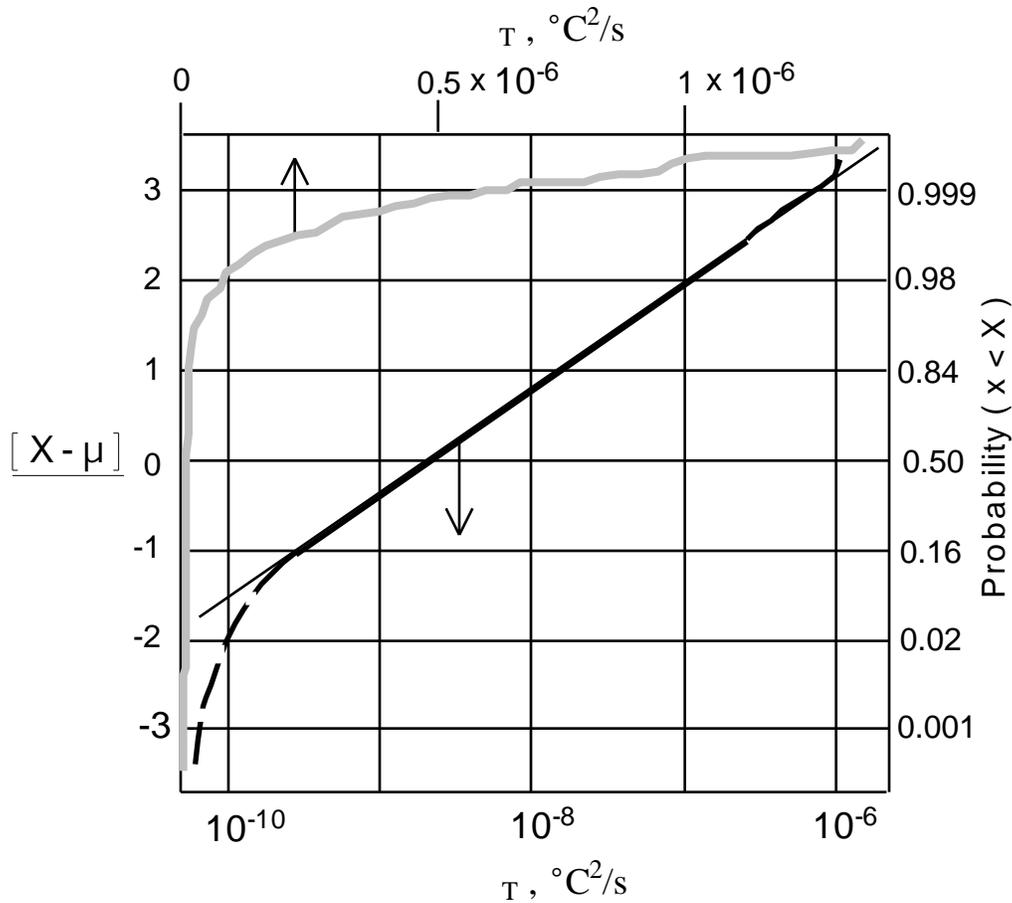

**Figure 7.** Temperature dissipation rate $\chi_T$ averaged over 4 m, compared to normal distribution (shaded line, upper abscissa) and lognormal distribution (dark line, lower abscissa). The data (7645 samples, over 30.8 km) were collected on 51st Cruise of AKADEMIK KURCHATOV from seasonal thermocline depths off California. The dashed portion of the lognormality plot may be affected by noise.

Fig. 7 shows a lognormality plot of the data (dark line, lower abscissa) that is very close to a straight line for a wide range of     values, indicating excellent agreement with a lognormal distribution function. The departure (dashed line) for values of $\chi_T$ less than $3 \times 10^{-10}$ °$C^2$ $s^{-1}$ is probably due to digitizing noise, rather than noise from the Artemyeva microconductivity sensor. From the slope of the straight line, $\sigma^2_{\ln \chi_T}$ is 3.9. The normality comparison (shaded line, upper abscissa) shows the samples are decidedly not normally distributed because the shaded line is so curved.

By averaging the data over different length scales r, and estimating $\sigma^2_{\ln \chi_T}$ values from plots such as Fig. 7, it is possible to estimate the universal intermittency constant $\mu$ that appears in hypothesis 4 of Table 1. Figure 8 shows a plot of $\sigma^2_{\ln \chi_T}$ versus r computed by Dr. Baker from the Soviet data for 11 different averaging lengths r ranging from 1 to 1024 m in octaves.

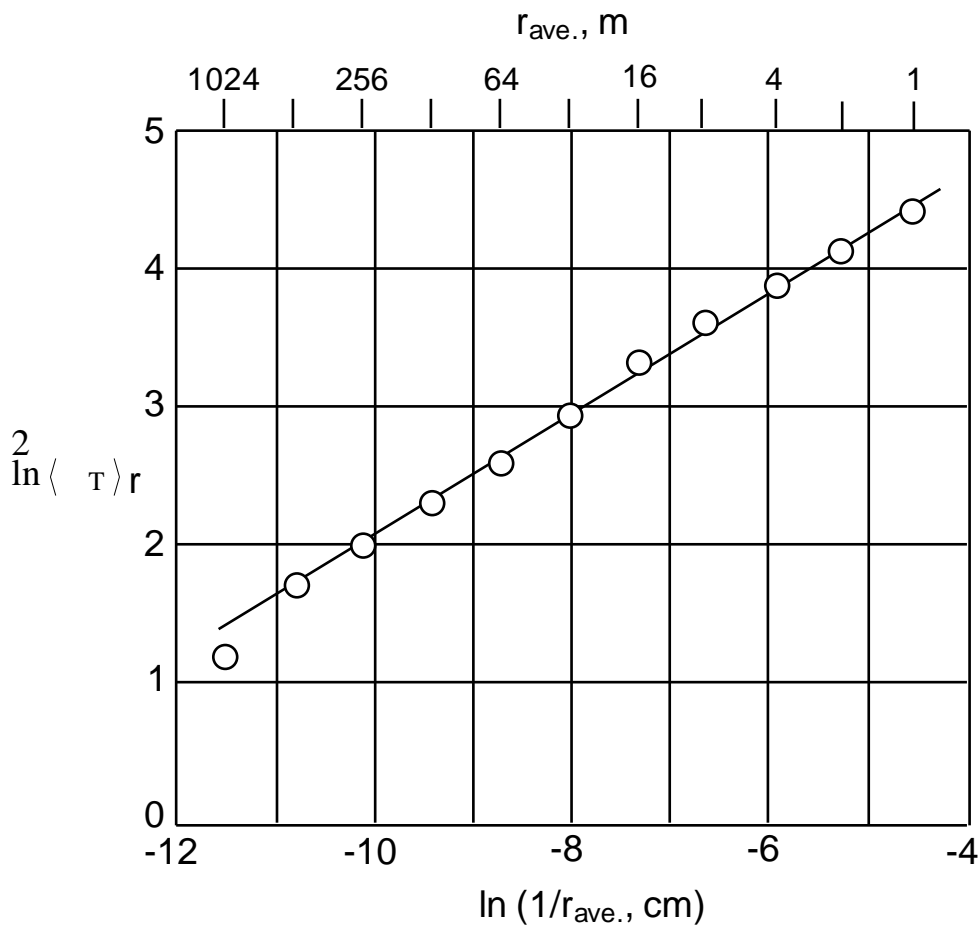

**Figure 8.** Intermittency factor versus ln (1/r), for averaging lengths r = 1-1024 m (circles), for the 30.8 km of data of Fig. 7. The slope of the line is the universal scalar intermittency constant $\mu$ by hypothesis 4 of Table 1, and is 0.44±0.01 for the fitted line.

The straight line formed by the data in Fig. 8 strongly supports the extension of the Kolmogorov intermittency hypothesis for scalars, hypothesis 4 of Table 1, and indicates that µ is 0.44±0.01. A similar plot for velocity dissipation rates in the marine atmospheric boundary layer by Gibson and Masiello (1971) gave µ = 0.47±0.03, and Gibson, Stegen and McConnell (1970) found µ 0.5±0.1 from the viscous dissipation rate spectral slope for similar data. From the Gurvich and Yaglom (1967) model, this equality of µ and µ suggests that the ratios of averaging scales are the same for both scalar and velocity dissipation rates for which averages over increasingly different length scales become statistically independent, and that the variance of the random variable formed from the ratio of such averages is the same for  as it is for  . That is,

$$\langle \ \rangle_r \text{ indep.} \langle \ \rangle_{r+kr} ; \langle \ \rangle_r \text{ indep.} \langle \ \rangle_{r+kr} ;$$
$$\text{var}(\langle \ \rangle_r / \langle \ \rangle_{r+kr}) \quad \text{var}(\langle \ \rangle_r / \langle \ \rangle_{r+kr}) \quad (10)$$

which is a sufficient, but not necessary, condition to give µ   µ. The dependence of this result on fossilization effects is presently unknown, but is probably small.

## 5. Results and Conclusions

Kolmogorov's universal similarity hypotheses for high Reynolds number turbulent velocity fields have been extended in Table 1 to passive scalar fields like temperature mixed by turbulence. The scalar hypotheses are complicated by the effects of Prandtl number. Some controversy exists for Pr values less than 0.02 that are difficult to test experimentally or simulate by computer. However, for the range 0.02 < Pr < 1000 the laboratory, field and computational evidence supports hypotheses 1ab, 2ab and 3ab. These assume a universal rate-of-strain mixing mechanism for the smallest scalar scales, diffusivity dependent rate-of-strain mixing at intermediate scales, and inertial-vortex stirring at larger scales at very high Reynolds, Froude and Rossby numbers.

The most useful applications of universal turbulence and mixing theories are for natural flows where Re, Fr and Ro are very large and the theories are most appropriate. Such flows are generally highly undersampled because of the large ranges of space and time, so the additional insight provided by universal similarity theory is particularly valuable. Evidence is reviewed that suggests the interstellar medium of the galaxy may be turbulent, and that fluctuations of scalar properties in the medium, like electron density and refractive index, behave like scalar fields mixed by turbulence.

The existence of spectra and structure functions with power law subranges consistent with those expected for turbulence and turbulent mixing does not prove flows with these properties are turbulent or that universal similarity of turbulence and turbulent mixing exists, but it is certainly strongly suggestive of these conclusions. Evidence of turbulence, and the Kolmogorovian similarity hypotheses, is most convincing when a wide variety of statistical parameters have been collected that are internally consistent for a particular flow, and mutually consistent for a variety of flows, with those expected from universal velocity and scalar mixing similarity.

On the other hand, only a single counterexample is needed to disprove the validity of universal similarity theories for turbulence or mixing. The agreement with universal velocity spectral forms of the Gargett, Osborn and Nasmyth (1984) Knight Inlet velocity data of Fig. 4 merely adds one more to hundreds of previous confirmations in all other turbulent flows tested. However, the strong departure of the temperature spectra from scalar universal forms for the same regions and flow regime shown in Fig. 5 demands an explanation or else universal scalar similarity theory is disproved, as claimed by Gargett (1985). The fossil turbulence and intermittency explanation given in §3 seems plausible, but a preferred scenario would be a repeat of the measurements in the same location taking the possibility of fossilization into account.

Neither extreme intermittency nor fossil turbulence effects have been taken into account in the interpretation of most oceanographic microstructure data sampled by western oceanographers. Consequently, the dropsonde sampling strategies adopted are subject to extreme quantitative, and even qualitative, undersampling errors. An example is shown in Fig. 6. Soviet measurements of turbulence and mixing using high frequency response velocity and temperature sensors on high speed towed bodies, combined with dropsonde sampling to reveal the physical processes in individual patches, seem better tailored to oceanic turbulence phenomena, and have revealed much higher intermittencies, and higher turbulence and mixing levels, than those usually inferred from dropsonde data sets alone.

The extremely intermittent and lognormal temperature dissipation rates and the logarithmic dependence of the scalar intermittency factor on the averaging length, shown in Fig.'s 7 and 8, seem to be strong evidence in support of the Kolmogorovian scalar intermittency hypothesis 4 of Table 1. Apparently the seasonal thermocline mixing in the California current is dominated by horizontal turbulence since the intermittency continues to decrease for averaging length scales of up to a kilometer. These interpretations are consistent with the towed body measurements of Washburn and Gibson (1984) in the seasonal thermocline of the North Pacific that also showed lognormality, and an intermittency factor $\sigma^2_{\ln}$ 4.8 close to those of Fig.'s 7 and 8. Effects of noise on these data are discussed by Baker and Gibson (1987).

It is interesting to note that the universal intermittency constants $\mu$ and $\mu$ for scalars and velocity are indistinguishable from each other, with values of about 0.45. The values of these constants reflect the range of scales required for averages of the dissipation rates to become statistically independent of averages on larger or smaller scales. Apparently the required scale ratio is the same for turbulent velocity dissipation as it is for scalar dissipation. Perhaps this should not be surprising, since the mixing and the geometry of the mixing regions that determine µ are caused by the turbulence and the geometry of the regions that determine µ.

## Acknowledgements

The author would like to acknowledge several useful conversations with colleagues in the preparation of this paper. Prof. Barney Rickett and Prof. Tom Phillips were very helpful in explaining elementary astronomical techniques. Particular thanks are due to Prof. Iosif Lozovatskiy, Dr. Valariy Nabatov, Dr. Anatoly Erofeev and Ms. Tamara Artemyeva for their kind assistance on the KURCHATOV. Director V. Yastrebov, Prof. Rostislav Ozmidov and Dr. V. Paka of the Shirshov Inst. of Oceanology, and the Academy of Sciences of the USSR, made it possible for the author and Dr. Mark Baker to join the AK51 Cruise. The data for Figures 7 and 8 were collected by Artemyeva, Lozovatskiy, Nabatov and others; Erofeev and Baker carried out the conversion from USSR to US formats on the ship; and Dr. Baker reduced the data and kindly provided the original plots. Prof. Owen Phillips sent a preprint of his contribution to this volume, which was very helpful.